\documentclass[showpacs,amsmath,amssymb,twocolumn,aps,pra,superscriptaddress]{revtex4-1}
\usepackage{hyperref}
\usepackage{verbatim}
\usepackage{amsmath}
\usepackage{latexsym}
\usepackage{revsymb}
\usepackage{yfonts}
\usepackage{ifthen}

\usepackage{natbib}
\usepackage{amsfonts}
\usepackage{amsmath}
\usepackage{amssymb}
\usepackage{amsthm}
\usepackage{graphicx}
\usepackage{bm}
\usepackage{bbm}
\usepackage{epsfig,color,amssymb}
\usepackage{subfigure}
\usepackage{amsfonts}
\usepackage{amscd}
\usepackage{amsmath}
\usepackage{multirow}
\usepackage{chemarrow}
\usepackage{dcolumn}
\usepackage{bm}
\usepackage{graphicx}
\usepackage{enumerate}
\usepackage{epsfig}
\usepackage{subfigure}
\usepackage{xcolor}
\usepackage{multirow}
\usepackage{ulem}
\usepackage{braket}
\usepackage{comment}
\usepackage{enumitem}
\usepackage{amsthm}
\renewcommand{\emph}[1]{\textit{#1}}

\begin{document}

\title{Coherent-State-Based Twin-Field Quantum Key Distribution}

\author{Hua-Lei Yin}\email{hlyin@nju.edu.cn}
\author{Zeng-Bing Chen}\email{zbchen@nju.edu.cn}
\affiliation{National Laboratory of Solid State Microstructures and School of Physics, Nanjing University, Nanjing 210093, China}


\begin{abstract}
Large-scale quantum communication networks are still a huge challenge due to the rate-distance limit of quantum key distribution (QKD). Recently, twin-field (TF) QKD has been proposed to overcome this limit.
Here, we prove that coherent-state-based TF-QKD is a time-reversed entanglement protocol, where the entanglement generation is realized with entanglement swapping operation via an entangled coherent state measurement. We propose a coherent-state-based TF-QKD with optimal secret key rate under symmetric and asymmetric channels by using coherent state and cat state coding. Furthermore, we show that our protocol can be converted to all recent coherent-state-based TF-QKD protocols by using our security proof. By using the entanglement purification with two-way classical communication, we improve the transmission distance of all coherent-state-based TF-QKD protocols.
\end{abstract}

\maketitle

Since the first quantum key distribution (QKD) experiment with 32 cm free-space channel~\cite{bennett1992experimental}, a lot of efforts have been devoted to achieving long-distance QKD.
Recently, the maximum distance of point-to-point QKD has been pushed up to 421 km ultralow-loss fiber~\cite{Boaron:2018:Secure}. Several experiments show that quantum-limited measurement~\cite{takenaka2017satellite,gunthner2017quantum} and QKD~\cite{liao2017satellite} can be demonstrated by using satellite-to-ground downlink with more than 1000 km free-space channel. Furthermore, measurement-device-independent (MDI) QKD~\cite{lo2012measurement} has been performed over 404 km ultralow-loss fiber~\cite{Yin:2016:Measurement} by using the optimal four-intensity set~\cite{zhou2016making}, which is immune to any attack on detection~\cite{braunstein2012side} by exploiting the Bell state measurement. Further increasing the fiber-based transmission distance without quantum repeater is a difficult obstacle to overcome. In the literature, without the help of trusted relay or quantum repeater, people believe that the limit is approximately 500 km fiber~\cite{gisin2015far}. The strong evidence comes from the secret key agreement capacity of repeaterless quantum channel~\cite{takeoka2014fundamental,pirandola2017fundamental}, where the optimal rate is linear scaling with the transmittance of two communication parties, known as the repeaterless bound~\cite{pirandola2017fundamental}.

A breakthrough called twin-field (TF) QKD~\cite{lucamarini2018overcoming} has been proposed to break this bound, resulting in many variants~\cite{ma2018phase,tamaki2018information,Wang2018Sending,yin2018practical,cui:2018:phase,curty:2018:simple,Lin:2018:A,primaatmaja2019versatile,xu2019general} and experimental demonstrations~\cite{minder2019experimental,liu2019experimental,wang2019beating,zhong2019proof}. However, each security proof of the coherent-state-based TF-QKD~\cite{ma2018phase,cui:2018:phase,curty:2018:simple,Lin:2018:A}, or called phase-matching QKD, is carefully tailored. Cat state, superposition of coherent states with two opposite phases, as an important resource, has been widely used for quantum information processing, including quantum computation~\cite{Lund:2008:Fault}, quantum teleportation~\cite{Andersen:2013:High}, quantum repeater~\cite{sangouard2010quantum,Brask:2010:Hybrid}, QKD~\cite{yin2014long} and quantum metrology~\cite{Joo:107:Quantum}. Importantly, cat states have been successfully generated and exploited to demonstrate various quantum tasks~\cite{ourjoumtsev2006generating,ourjoumtsev2007generation,vlastakis2015characterizing,ulanov2017quantum,le2018remote}.

Here, we point out the physics in coherent-state-based TF-QKD is exactly entanglement swapping operation via the entangled coherent state (ECS) measurement.
The coherent-state-based TF-QKD is a time-reversed entanglement protocol by using ECS measurement, which is similar with the MDI-QKD by using the Bell state measurement.
We propose a coherent-state-based TF-QKD protocol under symmetric and asymmetric channels by using coherent state and cat state coding.
The entanglement purification with one-way ~\cite{lo1999unconditional,Shor:2000:Simple} and two-way~\cite{gottesman2003proof} classical communication techniques are used to prove the security of our protocol against coherent attacks in the asymptotic regime.
The secret key rate of our protocol is larger than Refs.~\cite{ma2018phase,cui:2018:phase,curty:2018:simple}. Furthermore, we show that our protocol can be converted to other coherent-state-based protocols~\cite{ma2018phase,cui:2018:phase,curty:2018:simple,Lin:2018:A} by using our security proof, which means that all coherent-state-based TF-QKD can be unified under a single framework. We consider the TF-QKD with two-way classical communication, which significantly improves the transmission distance of all coherent-state-based TF-QKD protocols with large misalignment.

\section*{Results}

\noindent
\textbf{ECS measurement.}
Generally, the symmetric beam splitter (BS) and single-photon detectors are used to implement the interference measurement of TF-QKD~\cite{lucamarini2018overcoming}.
One can assume that the two inputs of BS are $a$ and $b$ modes while the two output modes are $\tilde{a}=(a+b)/\sqrt{2}$ and $\tilde{b}=(a-b)/\sqrt{2}$.
The four two-mode ECS forms~\cite{sanders1992entangled,jeong2002purification} are $\ket{\Phi^{\pm}}=\left(\ket{\alpha}\ket{\alpha}\pm\ket{-\alpha}\ket{-\alpha}\right)/\sqrt{N_{\pm}}$
and $\ket{\Psi^{\pm}}=\left(\ket{\alpha}\ket{-\alpha}\pm\ket{-\alpha}\ket{\alpha}\right)/\sqrt{N_{\pm}}$,
where $N_{\pm}=2(1\pm e^{-4\mu})$ are the normalization factors and $\mu=|\pm\alpha|^{2}$ is the intensity of coherent states $\ket{\pm\alpha}$. The four ECSs are sometimes called quasi-Bell states. The quantum states $\ket{\pm\alpha}$ constitute the quasi-computational basis while the quantum states $\ket{\xi^{\pm}(\alpha)}=(\ket{\alpha}\pm \ket{-\alpha})/\sqrt{2}$ constitute the quasi-dual basis.
After passing through the lossless symmetric BS, the four states become
\begin{equation}
\begin{aligned}\label{eq2}
\ket{\Phi^{+}}_{ab}& \xrightarrow{\textrm{BS}}\ket{\textrm{even}}_{\tilde{a}}\ket{0}_{\tilde{b}},~~~\ket{\Phi^{-}}_{ab} \xrightarrow{\textrm{BS}} \ket{\textrm{odd}}_{\tilde{a}}\ket{0}_{\tilde{b}},\\
\ket{\Psi^{+}}_{ab}& \xrightarrow{\textrm{BS}}\ket{0}_{\tilde{a}}\ket{\textrm{even}}_{\tilde{b}},~~~\ket{\Psi^{-}}_{ab} \xrightarrow{\textrm{BS}} \ket{0}_{\tilde{a}}\ket{\textrm{odd}}_{\tilde{b}},\\
\end{aligned}
\end{equation}
where $\ket{\textrm{even}}_{\tilde{a}}$ ($\ket{\textrm{odd}}_{\tilde{a}}$) means that the output mode $\tilde{a}$ contains even (odd) photon numbers.
If we consider the case of ideal photon-number-resolving detector and lossless channel, one can unambiguously discriminate the four ECSs by performing photon-number parity measurement.

For the case of lossy channel and threshold detector, one can only discriminate the case with or without detector clicks.
Generally, a successful detection event in TF-QKD~\cite{lucamarini2018overcoming} is defined that one and only one detector clicks. Therefore, we make only detector $L$ ($R$) clicking represent that the result of the ECS measurement is the state $\ket{\Phi^{-}}$ $\left(\ket{\Psi^{-}}\right)$. Due to decoherence of the cat states in lossy channel, the states $\ket{\Phi^{+}}$ and $\ket{\Psi^{+}}$ will always be mistakenly measured as quantum states $\ket{\Phi^{-}}$ and $\ket{\Psi^{-}}$, respectively. However, the corresponding probabilities can be restricted to be very low when the optical intensity is low and there is no eavesdropper's disturbance.
The post-selected joint quantum states of two legitimate users have quantum correlations, which then means that coherent-state-based TF-QKD have MDI characteristic.

\begin{figure}
\centering
\includegraphics[width=8cm]{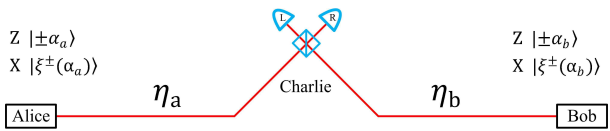}
\caption{The coherent-state-based TF-QKD with coherent state and cat state coding. Alice (Bob) randomly prepares coherent states $\ket{\pm\alpha_{a(b)}}$ and cat states $\ket{\xi^{\pm}(\alpha_{a(b)})}$ if choosing the $Z$ and $X$ bases, respectively. Alice and Bob use the insecure channels $\eta_{a}$ and $\eta_{b}$ to send the optical pulses to untrusted Charlie, who is supposed to perform the ECS measurement on the two incoming pulses. Only $L$ ($R$) detector click means a successful measurement outcome $\ket{\Phi^{-}}$ ($\ket{\Psi^{-}}$). The case of $\mu_{a}\eta_{a}=\mu_{b}\eta_{b}$ is required to keep perfect interference in the $Z$ basis, where intensity $\mu_{a(b)}=|\alpha_{a(b)}|^{2}$ and $\eta_{a(b)}$ is the efficiency between Alice (Bob) and Charlie.
} \label{f1}
\end{figure}

\bigskip
\noindent
\textbf{TF-QKD with cat state.}
We introduce a coherent-state-based TF-QKD with coherent state and cat state coding, as shown in Fig. \ref{f1}.
\emph{State preparation.} Alice (Bob) randomly chooses the $Z$ and $X$ bases with probabilities $p_{Z}$ and $p_{X}$. For the $Z$ basis, Alice (Bob) randomly prepares coherent state optical pulses $\ket{\alpha_{a(b)}}$ and $\ket{-\alpha_{a(b)}}$ with equal probabilities for the logic bits $0$ and $1$. For the $X$ basis, Alice (Bob) randomly prepares cat state optical pulses $\ket{\xi^{+}(\alpha_{a(b)})}$ and $\ket{\xi^{-}(\alpha_{a(b)})}$ with equal probabilities for the logic bits $0$ and $1$.
\emph{Entanglement measurement.} Alice and Bob send the optical pulses to the untrusted Charlie through insecure quantum channel with efficiency $\eta_{a}$ and $\eta_{b}$ (with detector efficiency taken into account). Charlie is supposed to perform the ECS measurement. For example, he let the two optical pulses interfere in the symmetric BS which would be detected by two threshold detectors $L$ and $R$.
\emph{Announcement.} Charlie publicly discloses whether he has obtained a successful measurement result and which ECS is acquired. Alice and Bob only keep the data of successful measurement and discard the rest.
\emph{Reconciliation.} Alice and Bob announce their bases over an authenticated classical channel. They only keep the data of the same basis and discard the rest. For the $Z$ basis, Bob flips his key bit if Charlie announces a result with $\ket{\Psi^{-}}$. For the $X$ basis, Bob always flips his key bit.
\emph{Parameter estimation.} The data of the $Z$ basis are used for constituting raw key and calculating the gain $Q_{Z}$ and quantum bit error rate (QBER) $E_{Z}$ of the $Z$ basis.
The data of the $X$ basis are all announced to calculate QBER $E_{X}$ of the $X$ basis.
\emph{Key distillation.} Alice and Bob exploit the error correction and privacy amplification to distill secret key.

\begin{figure*}
\centering
\includegraphics[width=15cm]{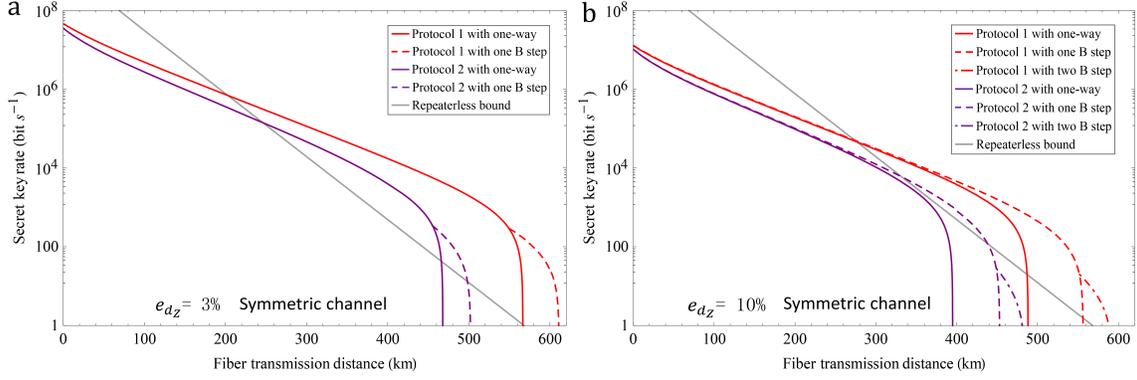}
\caption{The secret key rate under symmetric channel. \textbf{a (b)}, The misalignment of the $Z$ basis is $e_{d_{Z}}=3\%~(10\%)$. Protocols 1 and 2 denote our protocol and the protocol proposed in Refs.~\cite{cui:2018:phase,curty:2018:simple} with three-intensity phase-randomized coherent state, respectively.
We optimize the intensity of coherent state in the $Z$ basis for each transmission loss. The repeaterless bound~\cite{pirandola2017fundamental} is also shown in the figure.
} \label{f2}
\end{figure*}

Here, we prove that coherent-state-based TF-QKD with coherent state and cat state coding is secure against coherent attacks in the asymptotic regime.
Any one of two successful detections is enough for proving the security.
The coherent-state-based TF-QKD can be regarded as a time-reversed entanglement protocol, where Alice and Bob prepare maximally entangled state $\ket{\psi}_{a'a}=\big(\ket{+z}_{a'}\ket{\alpha_{a}}_{a(b)}+\ket{-z}_{a'}\ket{-\alpha_{a}}_{a}\big)/\sqrt{2}$ and $\ket{\psi}_{b'b}=\big(\ket{+z}_{b'}\ket{\alpha_{b}}_{b}+\ket{-z}_{b'}\ket{-\alpha_{b}}_{b}\big)/\sqrt{2}$, respectively. $\ket{\pm z}$ are the eigenstates of Pauli's $Z$ operator. Alice and Bob keep the qubit and send the optical mode to Charlie, who is supposed to perform entanglement swapping via the ECS measurement. Thereby, the bipartite states between Alice and Bob have quantum correlation through Charlie's entanglement swapping operation. One can use the entanglement purification technique~\cite{lo1999unconditional} to distill maximally entangled state and generate the secret key. If Alice and Bob measure qubits before sending optical pulses in the virtual entanglement protocol, it will become the prepare-and-measurement protocol, i.e., coherent-state-based TF-QKD. Details can be found in supplemental material. The efficient QKD scheme~\cite{lo2005efficient} can be directly applied, where we let $p_{Z}\approx1$ in the asymptotic limit.
The secret key rate of our coherent-state-based TF-QKD with one-way classical communication~\cite{Shor:2000:Simple} in the asymptotic limit is
\begin{equation}
\begin{aligned}\label{}
R=Q_{Z}[1-fh(E_{Z})-h(E_{X})],
\end{aligned}
\end{equation}
where $h(x)=-x\log_{2}x-(1-x)\log_{2}(1-x)$ is the Shannon entropy and $f=1.16$ is the error correction efficiency. In our simulation, without Charlie's disturbance, we have gain $Q_{Z}=(1-p_{d})[1-(1-2p_{d})e^{-2x}]$, QBERs $E_{Z}=(1-p_{d})[e_{d_{Z}}(1-e^{-2x})+p_{d}e^{-2x}]$ and $E_{X}=\frac{1}{2}\{1+e^{-2(\mu_{a}+\mu_{b})}[1-(1-2p_{d})e^{2x}]/[1-(1-2p_{d})e^{-2x}]\}$, where $x=\mu_{a}\eta_{a}=\mu_{b}\eta_{b}$, $e_{d_{Z}}$ is the misalignment rate of the $Z$ basis and $p_{d}$ is the dark count rate. The misalignment of the $X$ basis can be neglected since Bob always flips his bit.

Here, we exploit the two-way entanglement purification~\cite{gottesman2003proof} into our coherent-state-based TF-QKD protocol to increase transmission distance. Specifically, before implementing \emph{Key distillation} step, Alice randomly permutes all raw key bits and divides them into two groups, Bob does the same. Alice and Bob compute a parity on raw key of two groups and compare the parities. The second group is always discarded. If their parities are the same, they keep the bit of the first group. Otherwise, they discard it. One can repeat the above operation once for each B step.
After $k$th  B step is applied, the gain, bit and phase error rates can be given by
$Q_{Z}^{k}=\frac{1}{2}A^{k-1}Q_{Z}^{k-1}$, $E_{Z}^{k}=\left(E_{Z}^{k-1}\right)^{2}/A^{k-1}$, and
$E_{X}^{k}=2E_{X}^{k-1}\left(1-E_{Z}^{k-1}-E_{X}^{k-1}\right)/A^{k-1}$,
where we have $A^{k-1}=\left(1-E_{Z}^{k-1}\right)^{2}+\left(E_{Z}^{k-1}\right)^{2}$, $Q_{Z}^{0}=Q_{Z}$, $E_{Z}^{0}=E_{Z}$ and $E_{X}^{0}=E_{X}$.
The secret key rate of our coherent-state-based TF-QKD after $k$th B step in the asymptotic limit is
\begin{equation}
\begin{aligned}\label{}
R^{k}=Q_{Z}^{k}[1-fh(E_{Z}^{k})-h(E_{X}^{k})].
\end{aligned}
\end{equation}

\bigskip
\noindent
\textbf{Converting to other protocols.}
Without loss of generality, let positive-operator valued measure $E_{10}$ and $E_{01}$ ($E_{s}=E_{10}+E_{01}$) denote the successful measurement results with ECSs $\ket{\Phi^{-}}$ and $\ket{\Psi^{-}}$; let ${\rm \hat{P}}(\ket{u,v}):=\ket{u}\bra{u}\otimes\ket{v}\bra{v}$ with $\ket{u,v}=\ket{u}\ket{v}$.
The density matrix of the $Z$ and $X$ bases are $\rho_{Z}=\frac{1}{4}[{\rm \hat{P}}(\ket{\alpha_{a},\alpha_{b}})+{\rm \hat{P}}(\ket{\alpha_{a},-\alpha_{b}})+{\rm \hat{P}}(\ket{-\alpha_{a},\alpha_{b}})+{\rm \hat{P}}(\ket{-\alpha_{a},-\alpha_{b}})]$ and $\rho_{X}=\frac{1}{4}[{\rm \hat{P}}(\ket{\xi^{+}(\alpha_{a}),\xi^{+}(\alpha_{b})})+{\rm \hat{P}}(\ket{\xi^{+}(\alpha_{a}),\xi^{-}(\alpha_{b})})+{\rm \hat{P}}(\ket{\xi^{-}(\alpha_{a}),\xi^{+}(\alpha_{b})})+{\rm \hat{P}}(\ket{\xi^{-}(\alpha_{a}),\xi^{-}(\alpha_{b})})]$, where we have $\rho_{X}\equiv\rho_{Z}=\rho$.
For the cases of $\ket{\xi^{+}(\alpha_{a}),\xi^{+}(\alpha_{b})}$ and $\ket{\xi^{-}(\alpha_{a}),\xi^{-}(\alpha_{b})}$, they always generate the error gain since Bob always flips his bit, and the corresponding density matrix is
$\rho_{X}^{E}=\frac{1}{4}[{\rm \hat{P}}(\ket{\xi^{+}(\alpha_{a}),\xi^{+}(\alpha_{b})})+{\rm \hat{P}}(\ket{\xi^{-}(\alpha_{a}),\xi^{-}(\alpha_{b})})]$. Let $Q_{X}$ and $Q_{X}^{E}$ represent the gain and error gain of the $X$ basis.
In the case of asymptotic limit, we always have $Q_{X}\equiv Q_{Z}=\textrm{Tr}(\rho E_{s})$  and $Q_{X}^{E}=\textrm{Tr}(\rho_{X}^{E} E_{s})$.
Therefore, the QBER $E_{X}$ (phase error rate of the $Z$ basis) in the asymptotic limit can be given by
$E_{X}=Q_{X}^{E}/Q_{X}=Q_{X}^{E}/Q_{Z}$.
If one can acquire an upper bound of $Q_{X}^{E}$, the QBER $E_{X}$ can be bounded.

By using the entanglement purification with one-way~\cite{lo1999unconditional,Shor:2000:Simple} and two-way~\cite{gottesman2003proof} classical communication to prove security, we only require the estimation of the QBER $E_{X}$. This means that we do not need to prepare cat state if we can acquire the QBER $E_{X}$ through alternative method. The alternative method need to ensure that the prepared state by Alice (Bob) is linearly dependent, which cannot allow Charlie to implement unambiguous-state-discrimination attack~\cite{tang2013source} before performing the entanglement swapping. Here, we show that our protocol can be converted to the coherent-state-based protocols of Refs.~\cite{ma2018phase,cui:2018:phase,curty:2018:simple,Lin:2018:A} by using our security proof.

For the protocol proposed in Ref.~\cite{Lin:2018:A}, Alice and Bob randomly prepare coherent state $\ket{e^{i\theta_{a}}\sqrt{\mu_{a}}}$ and $\ket{e^{i\theta_{b}}\sqrt{\mu_{b}}}$ if they choose the $X$ basis, where they need phases $\theta_{a(b)}\in[0,2\pi)$ and infinite intensities $\mu_{a(b)}$. As pointed out in Ref.~\cite{Lin:2018:A}, the operator $\rho_{X}^{E}$ can be approximated to arbitrary precision in the Hilbert-Schmidt norm by the discrete diagonal coherent state representation
$\rho_{X}^{E}=\sum_{i=1}^{\infty}\lambda_{i}{\rm \hat{P}}(\ket{\omega_{a}^{i},\omega_{b}^{i}})$,
where $\ket{\omega_{a}^{i},\omega_{b}^{i}}$ is the tensor product of coherent state and $\lambda_{i}$ is complex number. Thereby, the error gain $Q_{X}^{E}$ can be precisely obtained by using coherent states with infinite intensities. Indeed, in the ideal situation with symmetric channel, $\mu=\mu_{a}=\mu_{b}$ and $\eta=\eta_{a}=\eta_{b}$, the secret key rate of this protocol by using our security proof with one-way classical communication is given by
$R=(1-e^{-\mu\eta})\left[1-h\left(\frac{1-e^{-4\mu+2\mu\eta}}{2}\right)\right]$,
which is the same with the results of Ref.~\cite{Lin:2018:A}.

For the protocol proposed in Refs.~\cite{cui:2018:phase,curty:2018:simple}, Alice and Bob randomly prepare phase-randomized coherent state if they choose the $X$ basis.
As pointed out in Ref~\cite{cui:2018:phase,curty:2018:simple}, by using the Cauchy-Schwarz inequality, we can bound the error gain $Q_{X}^{E}$ with photon-number state, i.e., $Q_{X}^{E}=\textrm{Tr}(\rho_{X}^{E} E_{s})\leq \left(\sum_{n,m=0}^{\infty}\sqrt{P_{2n}^{a}P_{2m}^{b}Y_{2n,2m}}\right)^{2}+\left(\sum_{n,m=0}^{\infty}\sqrt{P_{2n+1}^{a}P_{2m+1}^{b}Y_{2n+1,2m+1}}\right)^{2}$. $Y_{n,m}$ is the yield given that Alice and Bob send $n$ and $m$ photon states and $P_{n}^{a(b)}=e^{-\mu_{a(b)}}\mu_{a(b)}^{n}/n!$. Decoy-state method~\cite{Hwang:2003:Quantum,wang2005beating,lo2005decoy} can be used to estimate the yield $Y_{n,m}$, which has been realized with finite intensities~\cite{cui:2018:phase,grasselli2019practical}. Here, we use the three-intensity, $0<\omega<\nu$, to estimate the yield, which can be found in Methods.

For the protocol proposed in Ref.~\cite{ma2018phase}, Alice and Bob always prepare the coherent state $\ket{e^{i(\theta_{a}+\kappa_{a}\pi)}\sqrt{\mu_{a}}}$ and $\ket{e^{i(\theta_{b}+\kappa_{b}\pi)}\sqrt{\mu_{b}}}$, where $\kappa_{a(b)}\in\{0,1\}$ and $\theta_{a(b)}\in[0,2\pi)$. They keep the raw key bit only if $|\theta_{a}-\theta_{b}|=0$ or $\pi$. As pointed out in Ref~\cite{ma2018phase}, one can introduce a virtual trusted party who prepares a state, splits it using symmetric BS, and sends it to both Alice and Bob. We have the following observations
${\rm \hat{P}}(\ket{\alpha})+{\rm \hat{P}}(\ket{-\alpha})={\rm \hat{P}}(\ket{\xi^{+}(\alpha)})+{\rm \hat{P}}(\ket{\xi^{-}(\alpha)})$ and ${\rm \hat{P}}(\ket{\xi^{+}(\sqrt{2}\alpha),0})+{\rm \hat{P}}(\ket{0,\xi^{+}(\sqrt{2}\alpha)})$$\xrightarrow{\textrm{BS}}$${\rm \hat{P}}(\ket{\xi^{+}(\alpha),\xi^{+}(\alpha)})$$+{\rm \hat{P}}(\ket{\xi^{-}(\alpha),\xi^{-}(\alpha)})$.
For the post-selected phase-matching, the error gain can be given by $Q_{X}^{E}=\sum_{n=0}^{\infty}e^{-2\mu}(2\mu)^{2n}Y_{2n}/(2n)!$, where we need to assume $\mu_{a}=\mu_{b}=\mu$. $Y_{n}$ is the yield given that the total photon number sent by Alice and Bob is $n$. Only even photon numbers have contribution to the phase error rate in our security proof which is only the same with the results of Ref.~\cite{ma2018phase} in the ideal situation.
Different from protocols of ours and Refs.~\cite{cui:2018:phase,curty:2018:simple,Lin:2018:A}, the protocol of Ref.~\cite{ma2018phase} seems to be only suitable for symmetric channel.

\begin{figure}
\centering
\includegraphics[width=8cm]{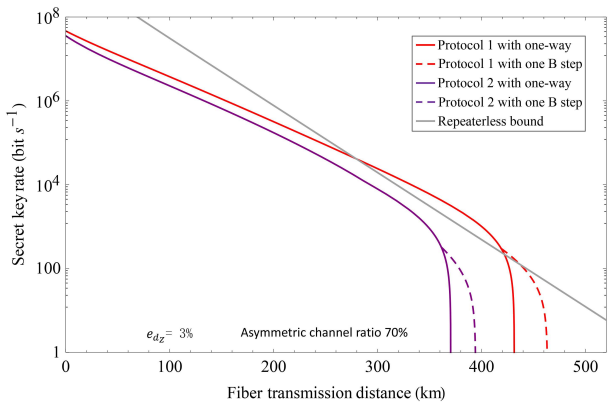}
\caption{The secret key rate under asymmetric channel. We optimize the intensity for each transmission loss. The repeaterless bound~\cite{pirandola2017fundamental} is also shown in the figure.
} \label{f3}
\end{figure}

\section*{Discussion}
For simulation, we use the following parameters. The inherent loss of fiber is $0.16$ dB/km, the efficiency and dark count rate of threshold single-photon detector are $\eta_{d}=85\%$ and $p_{d}=10^{-7}$.
For simplicity, we let Protocol 1 represent our coherent state and cat state coding protocol. Let Protocol 2 represent the protocol in Ref.~\cite{cui:2018:phase,curty:2018:simple} with three-intensity phase-randomized coherent state. Here, we fix the intensities of phase-randomized coherent state with $\nu=0.1$ and $\omega=0.02$. The secret key rate of our protocol is equal to that of protocol in Ref.~\cite{Lin:2018:A} since cat state can be approximated to arbitrary precision in the Hilbert-Schmidt norm by the discrete diagonal coherent state representation~\cite{Lin:2018:A}.
The performance of Protocols 1 and 2 under symmetric channel have been shown in Fig. \ref{f2}, which assumes 1 GHz system repetition rate~\cite{minder2019experimental}.
Here, we do not consider the performance of protocols in Ref.~\cite{ma2018phase}, whose secret key rate has been shown lower than protocols in Refs.~\cite{cui:2018:phase,curty:2018:simple,Lin:2018:A}.
The secret key rate and transmission distance of Protocol 1 are both superior to Protocol 2. The transmission distance of Protocols 1 and 2 can both be improved by using the two-way classical communication. Especially, the advantages are very clear when the system misalignment rate is large, like $e_{d_{Z}}=10\%$. The large system misalignment of TF-QKD is reasonable since the phase-locking and long-distance phase stabilization techniques in field are still difficult even with some experimental progresses~\cite{minder2019experimental,liu2019experimental,wang2019beating,zhong2019proof}. The performance of Protocols 1 and 2 under asymmetric channel have been shown in Fig. \ref{f3}. The secret key rate of Protocol 1 can still surpass the repeaterless bound when the asymmetric channel ratio is $70\%$. The asymmetric channel ratio is the ratio between A-C and A-B, where A-C (B) represents the distance between Alice and Charlie (Bob). Compare Figs. \ref{f2} with \ref{f3}, the performance of coherent-state-based TF-QKD is the best under symmetric channel due to the single-photon-type interference.

In summary, we propose a coherent-state-based TF-QKD with optimal secret key rate by using coherent state and cat state coding. By using the ECS measurement as the entanglement swapping operation, we unify all known coherent-state-based TF-QKD protocols under a single framework.
We have proved that the coherent-state-based TF-QKD is a time-reversed entanglement protocol, which means that one can use the known techniques of qubit-based QKD to further develop the coherent-state-based TF-QKD.
The results show that coherent-state-based TF-QKD is suitable for building quantum communication networks within hundreds of kilometers without trusted relay or quantum repeater. We remark that cat states used for high-speed QKD will have difficulty under current experimental conditions. However, current experiments on the demonstration of various quantum tasks with cat states are very active~\cite{ourjoumtsev2006generating,ourjoumtsev2007generation,vlastakis2015characterizing,ulanov2017quantum,le2018remote}, implying that cat states may become a practical resource with the rapid development of technology.

\section*{Methods}
\noindent
\textbf{Decoy-state analysis.}\label{ent}
The phase-randomized coherent state can be seen as a mixture of Fock states.  Let $Q_{a,b}$ represent the gain when Alice and Bob send phase-randomized coherent state with intensity $a$ and $b$, respectively. Let $Y_{n,m}$ represent the yield when Alice and Bob send $n$-photon and $m$-photon. Thereby, we have
$Q_{a,b}=\sum_{n=0}^{\infty}\sum_{m=0}^{\infty}e^{-a-b}\frac{a^{n}b^{m}}{n!m!}Y_{n,m}$.
Here, we exploit the decoy-state method with three-intensity to estimate the upper bound of the yield $Y_{n,m}$ with analytical method. The upper bound of $Y_{1,1}$, $Y_{0,2}$ and $Y_{2,0}$ can be given by
\begin{equation}
\begin{aligned}\label{}
Y_{1,1}&\leq \frac{e^{2\omega}Q_{\omega,\omega}-e^{\omega}(Q_{\omega,0}+Q_{0,\omega})+Q_{0,0}}{\omega^{2}},\\
\end{aligned}
\end{equation}
\begin{equation}
\begin{aligned}\label{}
Y_{0,2}&\leq \frac{\omega e^{\nu}Q_{0,\nu}-\nu e^{\omega}Q_{0,\omega}+(\nu-\omega)Q_{0,0}}{\nu\omega(\nu-\omega)/2},\\
\end{aligned}
\end{equation}
and
\begin{equation}
\begin{aligned}\label{}
Y_{2,0}&\leq \frac{\omega e^{\nu}Q_{\nu,0}-\nu e^{\omega}Q_{\omega,0}+(\nu-\omega)Q_{0,0}}{\nu\omega(\nu-\omega)/2},\\
\end{aligned}
\end{equation}
where we have $Y_{0,0}=Q_{0,0}$.
The upper bound of $Y_{0,n}$ and $Y_{n,0}$ with $n\geq3$ can be written as
\begin{equation}
\begin{aligned}\label{}
Y_{0,n}\leq {\rm min}\left\{1,~\frac{\omega e^{\nu}Q_{0,\nu}-\nu e^{\omega}Q_{0,\omega}+(\nu-\omega)Q_{0,0}}{\nu\omega(\nu^{n-1}-\omega^{n-1})/n!}\right\},\\
\end{aligned}
\end{equation}
and
\begin{equation}
\begin{aligned}\label{}
Y_{n,0}\leq {\rm min}\left\{1,~\frac{\omega e^{\nu}Q_{\nu,0}-\nu e^{\omega}Q_{\omega,0}+(\nu-\omega)Q_{0,0}}{\nu\omega(\nu^{n-1}-\omega^{n-1})/n!}\right\}.\\
\end{aligned}
\end{equation}
Let $F_{x,y}=e^{x+y}Q_{x,y}-e^{x}Q_{x,0}-e^{y}Q_{0,y}+Q_{0,0}$, the upper bound of $Y_{1,n}$ and $Y_{n,1}$ with $n\geq2$ can be given by
\begin{equation}
\begin{aligned}\label{}
Y_{1,n}\leq {\rm min}\left\{1,~\frac{\omega F_{\omega,\nu}-\nu F_{\omega,\omega}}{(\nu^{n}\omega^{2}-\nu\omega^{n+1})/n!}\right\},\\
\end{aligned}
\end{equation}
and
\begin{equation}
\begin{aligned}\label{}
Y_{n,1}\leq {\rm min}\left\{1,~\frac{\omega F_{\nu,\omega}-\nu F_{\omega,\omega}}{(\nu^{n}\omega^{2}-\nu\omega^{n+1})/n!}\right\}.\\
\end{aligned}
\end{equation}
Similarly, the upper bound of $Y_{n,m}$ with $n,m\geq2$ can be given by
\begin{equation}
\begin{aligned}\label{}
Y_{n,m}\leq &{\rm min}\Big\{1,\frac{\omega^{2}F_{\nu,\nu}-\nu\omega(F_{\nu,\omega}+F_{\omega,\nu})+\nu^{2}F_{\omega,\omega}}{\nu^{2}\omega^{2}(\nu^{n-1}-\omega^{n-1})(\nu^{m-1}-\omega^{m-1})/n!m!}     \Big\}.
\end{aligned}
\end{equation}

\noindent\textbf{Acknowledgments}\\
We gratefully acknowledges support from the National Natural Science Foundation of China under Grant No. 61801420 and Nanjing University.

\noindent\textbf{Author Contributions}\\
H.-L.Y. and Z.-B.C. conceived and designed the study. H.-L.Y. performed the numerical simulation.
All authors contributed extensively to the work presented in this paper.

\noindent\textbf{Additional Information}\\
Competing interests: The authors declare no competing interests.

\bibliographystyle{naturemag}



\newpage
\onecolumngrid

\maketitle
\section*{Supplementary Information 1: Heralded entanglement generation}

The four two-mode entangled coherent states (ECSs) \cite{sanders1992entangled1} can be written in different forms
\begin{equation}
\begin{aligned}\label{}
\ket{\Phi^{\pm}}&=\frac{1}{\sqrt{N_{\pm}}}\left(\ket{\alpha}\ket{\alpha}\pm\ket{-\alpha}\ket{-\alpha}\right)=\frac{1}{\sqrt{N_{\pm}}}\left(\ket{\xi^{+}(\alpha)}\ket{\xi^{\pm}(\alpha)}+\ket{\xi^{-}(\alpha)}\ket{\xi^{\mp}(\alpha)}\right)\\
&=\frac{1}{\sqrt{N_{\pm}}}\left(\ket{\xi^{+i}(\alpha)}\ket{\xi^{\mp i}(\alpha)}+\ket{\xi^{-i}(\alpha)}\ket{\xi^{\pm i}(\alpha)}\right),\\
\ket{\Psi^{\pm}}&=\frac{1}{\sqrt{N_{\pm}}}\left(\ket{\alpha}\ket{-\alpha}\pm\ket{-\alpha}\ket{\alpha}\right)=\frac{1}{\sqrt{N_{\pm}}}\left(\ket{\xi^{\pm}(\alpha)}\ket{\xi^{+}(\alpha)}-\ket{\xi^{\mp}(\alpha)}\ket{\xi^{-}(\alpha)}\right)\\
&=\frac{1}{i\sqrt{N_{\pm}}}\left(\ket{\xi^{\pm i}(\alpha)}\ket{\xi^{+i}(\alpha)}-\ket{\xi^{\mp i}(\alpha)}\ket{\xi^{-i}(\alpha)}\right),
\end{aligned}
\end{equation}
where the parameters $N_{\pm}=2(1\pm e^{-4\mu})$ are the normalization factors. The quantum states $\ket{\pm\alpha}$ are the coherent states containing $\mu=|\alpha|^2$ photons on average. The quantum states $\ket{\xi^{\pm}(\alpha)}=(\ket{\alpha}\pm\ket{-\alpha})/\sqrt{2}$ and $\ket{\xi^{\pm i}(\alpha)}=(\ket{\alpha}\pm i\ket{-\alpha})/\sqrt{2}$ are the non-normalized single-mode cat states. Considering a lossless and symmetric beam splitter (BS), the evolution of four ECSs after passing through the BS can be given by
\begin{equation}
\begin{aligned}\label{}
\ket{\Phi^{+}}_{ab}&\autorightarrow{\textrm{BS}}{\textrm{}}\frac{2e^{-\mu}}{\sqrt{N_{+}}}\sum_{n=0}^{\infty}\frac{(\sqrt{2}\alpha)^{2n}}{\sqrt{(2n)!}}\ket{2n}_{\tilde{a}}\ket{0}_{\tilde{b}}\Longrightarrow\ket{\textrm{even}}_{\tilde{a}}\ket{0}_{\tilde{b}},\\
\ket{\Phi^{-}}_{ab}&\autorightarrow{\textrm{BS}}{\textrm{}}\frac{2e^{-\mu}}{\sqrt{N_{-}}}\sum_{n=0}^{\infty}\frac{(\sqrt{2}\alpha)^{2n+1}}{\sqrt{(2n+1)!}}\ket{2n+1}_{\tilde{a}}\ket{0}_{\tilde{b}}\Longrightarrow\ket{\textrm{odd}}_{\tilde{a}}\ket{0}_{\tilde{b}},\\
\ket{\Psi^{+}}_{ab}&\autorightarrow{\textrm{BS}}{\textrm{}}\frac{2e^{-\mu}}{\sqrt{N_{+}}}\sum_{n=0}^{\infty}\frac{(\sqrt{2}\alpha)^{2n}}{\sqrt{(2n)!}}\ket{0}_{\tilde{a}}\ket{2n}_{\tilde{b}}\Longrightarrow\ket{\textrm{0}}_{\tilde{a}}\ket{\textrm{even}}_{\tilde{b}},\\
\ket{\Psi^{-}}_{ab}&\autorightarrow{\textrm{BS}}{\textrm{}}\frac{2e^{-\mu}}{\sqrt{N_{-}}}\sum_{n=0}^{\infty}\frac{(\sqrt{2}\alpha)^{2n+1}}{\sqrt{(2n+1)!}}\ket{0}_{\tilde{a}}\ket{2n+1}_{\tilde{b}}\Longrightarrow\ket{\textrm{0}}_{\tilde{a}}\ket{\textrm{odd}}_{\tilde{b}}.
\\
\end{aligned}
\end{equation}

In the virtual entanglement-based protocol, the entangled state prepared by Alice can be written as
\begin{equation}
\begin{aligned}\label{}
\ket{\psi}_{a'a}&=\frac{1}{\sqrt{2}}\left(\ket{+z}_{a'}\ket{\alpha}_{a}+\ket{-z}_{a'}\ket{-\alpha}_{a}\right)\\
&=\frac{1}{\sqrt{2}}\left(\ket{+x}_{a'}\ket{\xi^{+}(\alpha)}_{a}+\ket{-x}_{a'}\ket{\xi^{-}(\alpha)}_{a}\right)\\
&=\frac{1}{\sqrt{2}}\left(\ket{+y}_{a'}\ket{\xi^{-i}(\alpha)}_{a}+\ket{-y}_{a'}\ket{\xi^{+i}(\alpha)}_{a}\right),\\
\end{aligned}
\end{equation}
and the entangled state prepared by Bob can be written as
\begin{equation}
\begin{aligned}\label{}
\ket{\psi}_{b'b}&=\frac{1}{\sqrt{2}}\left(\ket{+z}_{b'}\ket{\alpha}_{b}+\ket{-z}_{b'}\ket{-\alpha}_{b}\right)\\
&=\frac{1}{\sqrt{2}}\left(\ket{+x}_{b'}\ket{\xi^{+}(\alpha)}_{b}+\ket{-x}_{b'}\ket{\xi^{-}(\alpha)}_{b}\right)\\
&=\frac{1}{\sqrt{2}}\left(\ket{+y}_{b'}\ket{\xi^{-i}(\alpha)}_{b}+\ket{-y}_{b'}\ket{\xi^{+i}(\alpha)}_{b}\right),\\
\end{aligned}
\end{equation}
where qubit states $\ket{\pm z}$, $\ket{\pm x}$ and $\ket{\pm y}$ are the eigenstates of Pauli's $Z$, $X$ and $Y$ operators. The bipartite qubit entanglement states $\rho_{a'b'}$ between Alice and Bob are generated by using the event-ready detection to implement upon the flying optical pulses, called entanglement swapping. Once Alice and Bob share qubit entanglement states $\rho_{a'b'}$ even with noise, they can exploit most previous security proof techniques to obtain secret key. Here, we use the entanglement purification techniques \cite{lo1999unconditional1,Shor:2000:Simple1,gottesman2003proof1} to prove the security of our protocols against coherent attacks in the asymptotic regime.

\section*{Supplementary Information 2: Entanglement purification and security proof of quantum key distribution}
Here we review the entanglement distillation protocol (EDP) of bipartite qubit systems and its relation with the security proof of quantum key distribution (QKD).
In the work of Bennett, Divincenzo, smolin and Wooters (BDSW) \cite{Bennett:1996:Mixed1}, it was shown that any bipartite qubit system density matrix can always be transformed into a diagonal form by local operations and classical communication. The diagonal forms of density matrix are in the Bell states:
\begin{equation}
\begin{aligned}\label{}
\ket{\psi_{1}}=\frac{1}{\sqrt{2}}(\ket{+z}\ket{+z}+\ket{-z}\ket{-z}),\\
\ket{\psi_{2}}=\frac{1}{\sqrt{2}}(\ket{+z}\ket{+z}-\ket{-z}\ket{-z}),\\
\ket{\psi_{3}}=\frac{1}{\sqrt{2}}(\ket{+z}\ket{-z}+\ket{-z}\ket{+z}),\\
\ket{\psi_{4}}=\frac{1}{\sqrt{2}}(\ket{+z}\ket{-z}-\ket{-z}\ket{+z}).\\
\end{aligned}
\end{equation}
By using the argument of BDSW \cite{Bennett:1996:Mixed1}, the density matrix $\rho$ describing Alice and Bob's qubit systems can be regarded as a classical mixture of the Bell states
\begin{equation}
\begin{aligned}\label{}
\rho=\lambda_{1}\ket{\psi_{1}}\bra{\psi_{1}}+\lambda_{2}\ket{\psi_{2}}\bra{\psi_{2}}+\lambda_{3}\ket{\psi_{3}}\bra{\psi_{3}}+\lambda_{4}\ket{\psi_{4}}\bra{\psi_{4}},
\end{aligned}
\end{equation}
normalized with $\sum_{i=1}^{4}\lambda_{i}=1$. If we let $\ket{\psi_{1}}$ be the reference state, the parameters $\lambda_{1}$, $\lambda_{2}$, $\lambda_{3}$ and $\lambda_{4}$ represent the probabilities of applying the Pauli $I$, $Z$, $X$ and $Y$ operators to either one of the qubit of the bipartite systems. Therefore, the parameters $\lambda_{1}$, $\lambda_{2}$, $\lambda_{3}$, $\lambda_{4}$ are the probabilities of no error, only phase flip error, only bit flip error, both bit and phase flip errors, respectively.
The hashing method and recurrence method have been proposed to implement the EDP in the BDSW argument \cite{Bennett:1996:Mixed1} if the density matrix is Bell-diagonal.
The job of EDP is to distill almost perfect Einstein-Podolsky-Rosen (EPR) pairs from the shared noise EPR pairs by using the local operations and classical communication to correct the bit and phase errors.

Due to the monogamy of entanglement, the eavesdropper's system almost has no quantum correlation with the system shared by Alice and Bob if they share nearly perfect pure EPR pairs. Therefore, Alice and Bob can measure the EPR pairs with the same basis to acquire the secret key while the leaked information is negligible. An important conclusion obtained in the Lo-Chau security proof \cite{lo1999unconditional1} is that the general state (highly entangled between different pairs) brings no advantage over a mixture of products of Bell states for the eavesdropper. It successfully reduces the quantum (joint) coherent attack to classical collective attack, which means that eavesdropper's probability of cheating successfully is negligible and the extracted secret key of QKD is secure against all possible attacks by using the EDP. A drawback of the Lo-Chau security proof is the requirement of quantum computer to implement the quantum error correction (bit and phase errors). The distillation rate of EPR pairs with one-way EDP \cite{Bennett:1996:Mixed1} in the asymptotic limit is
\begin{equation}
\begin{aligned}\label{}
r=1-h(e_{b})-H(e_{p}|e_{b}),
\end{aligned}
\end{equation}
where $h(x)=-x\log_{2}x-(1-x)\log_{2}(1-x)$ is the Shannon entropy.
The conditional Shannon entropy $H(e_{p}|e_{b})$ is given by \cite{yin2016security1}
\begin{equation} \label{mutual information}
\begin{aligned}
H(e_{p}|e_{b})=&-(1+a-e_{b}-e_{p})\log_{2}\frac{1+a-e_{b}-e_{p}}{1-e_{b}}\\
&-(e_{p}-a)\log_{2}\frac{e_{p}-a}{1-e_{b}}-(e_{b}-a)\log_{2}\frac{e_{b}-a}{e_{b}}-a\log_{2}\frac{a}{e_{b}}.
\end{aligned}
\end{equation}
where $e_{b}=\lambda_{3}+\lambda_{4}$ is bit error rate , $e_{p}=\lambda_{2}+\lambda_{4}$ is phase error rate and $a=\lambda_{4}$ quantifies the mutual information between bit and phase errors. If the parameter $a=e_{b}e_{p}$, one has $H(e_{p}|e_{b})=h(e_{p})$, which indicates no mutual information between bit and phase errors.

The entanglement-based QKD can be reduced to prepare-and-measure protocol by exploiting the Calderbank-Shor-Steane (CSS) error correction code in the Shor-Preskill security proof \cite{Shor:2000:Simple1}. One can decouple the phase error correction from the bit error correction in the CSS error correction code. Once Alice and Bob estimate the bit and phase error rates, they can choose appropriate CSS code to correct all the bit and phase errors. The phase error rate estimation method is arbitrary (direct measurement in the $X$ basis is not necessary).
The final measurement, such as the $Z$ basis, can be moved to the beginning since the $Z$ measurement commutes with other steps if we remove the phase error correction. Therefore, the quantum bit error correction can be replaced by classical bit error correction while the quantum phase error correction can be replaced by classical privacy amplification.
For the BB84 encoding \cite{bennett1984quantum1} with the $Z$ and $X$ bases, the secret key rate of the $Z$ basis with one-way classical communication in the Shor-Preskill security proof \cite{Shor:2000:Simple1} is
\begin{equation}
\begin{aligned}\label{}
r_{\textrm{BB84}}=1-h(e_{z})-h(e_{x}),
\end{aligned}
\end{equation}
where $e_{z}=e_{b}$ and $e_{x}=e_{p}$ are the quantum bit error rates (QBERs) of the $Z$ and $X$ bases. The parameter $a$ can be set to $e_{b}e_{p}$ in the BB84 encoding since there is no restriction on $a$ $\left(0\leq a\leq \min(e_{b},e_{p})\right)$, which means that there is no mutual information for the worst-case scenario. The six-state \cite{Bruss:1998:Optimal1} encoding QKD with one-way classical communication is proved by Lo \cite{lo2001proof1}, the corresponding secret key rate of the $Z$ basis is
\begin{equation}
\begin{aligned}\label{}
r_{\textrm{six-state}}=1-h(e_{z})-H(e_{x}|e_{z}),
\end{aligned}
\end{equation}
where mutual information parameter $a=(e_{z}+e_{x}-e_{y})/2$ exploiting the  QBER of the $Y$ basis is $e_{y}=\lambda_{2}+\lambda_{3}$. One can acquire the mutual information by using the extra $Y$ basis which means that the tolerant noise of six-state encoding is higher than the BB84 encoding.

Compared with the one-way EDP, the two-way EDP proposed by Gottesman and Lo \cite{gottesman2003proof1} has shown an advantage in tolerating noise. Except for the final random hashing used in one-way EDP, there are another two types of steps, B step and P step, in the Gottesman-Lo security proof \cite{gottesman2003proof1}. The B and P steps are used for decreasing the bit and phase error rates, respectively. Then the key can be extracted by applying random hashing. This is the reason why Gottesman-Lo's two-way EDP is able to tolerate more noise.

\emph{Definition of B step.} Alice and Bob perform a bilateral XOR operation on two EPR pairs and compare the measurement results of target pairs in the $Z$ basis after they randomly permute all the EPR pairs and divide them into two EPR pairs, control pairs and target pairs. The bilateral XOR measurement is used to detect the single bit error. It means that the measurement result is the same (different) given that the two EPR pairs have no bit error or both have a bit error (only one of the two EPR pairs has bit error). If the measurement outcomes are the same, they keep the control qubit; otherwise, they discard it. The B step requires two-way classical communication to change information between Alice and Bob. The B step is compatible with the prepare-and-measure protocol since the bilateral XOR operation of B step is equivalent to two measurement of $Z\otimes Z$.
If we assume that the noise EPR pairs are characterized by $\{e_{b},e_{p},a\}$, the new state is characterized by $\{\tilde{e}_{b},\tilde{e}_{p},\tilde{a}\}$ \cite{gottesman2003proof1} after one B step is applied,
\begin{equation}
\begin{aligned}\label{}
\tilde{e}_{b}=&\frac{e_{b}^{2}}{(1-e_{b})^{2}+e_{b}^{2}},\\
\tilde{e}_{p}=&\frac{2(1-e_{b}-e_{p}+a)(e_{p}-a)+2a(e_{b}-a)}{(1-e_{b})^{2}+e_{b}^{2}},\\
\tilde{a}=&\frac{2a(e_{b}-a)}{(1-e_{b})^{2}+e_{b}^{2}},\\
\end{aligned}
\end{equation}
where $p_{\textrm{B}}^{s}=[(1-e_{b})^{2}+e_{b}^{2}]/2$ is the probability of survival EPR pairs after one B step. The factor $1/2$ stems from the fact that only half of the initial EPR pairs are control pairs.
For the BB84 encoding, $a$ is a freedom parameter $0\leq a\leq \min(e_{b},e_{p})$, the worst case of B or P steps is $a=0$ in the two-way EDP proved by Gottesman and Lo \cite{gottesman2003proof1}, which is different from the one-way EDP \cite{Shor:2000:Simple1} with $a=e_{b}e_{p}$.

\emph{Definition of P step.} Alice and Bob randomly permute all EPR pairs and divide them into three groups, one target and two control EPR pairs.
They perform two bilateral XOR on three EPR pairs by one target and two control pairs. By measuring the two control pairs in the $X$ basis and comparing the measurement results, they can find the phase error syndrome. However, the phase error cannot be detected and corrected in the prepare-and-measure protocol. The P step is reduced to implement the classical XOR operation among the three bits to generate one bit in the prepare-and-measure protocol if the $Z$ basis measurement is performed before the P step. Therefore, if we assume that the noise EPR pairs are characterized by $\{e_{b},e_{p},a\}$, the new EPR pairs are characterized by $\{\tilde{e}_{b},\tilde{e}_{p},\tilde{a}\}$ \cite{gottesman2003proof1} after one P step is implemented,
\begin{equation}
\begin{aligned}\label{}
\tilde{e}_{b}&=3e_{b}(1-e_{b})^{2}+e_{b}^{3},\\
\tilde{e}_{p}&=3e_{p}^{2}(1-e_{p})+e_{p}^{3},\\
\tilde{a}&=3a(e_{p}-a)(2-2e_{b}-e_{p}+a)+3(e_{b}-a)[a^{2}+(e_{p}-a)^{2}]+a^{3},\\
\end{aligned}
\end{equation}
where $p_{\textrm{P}}^{s}=1/3$ is the probability of survival EPR pairs after one P step since only one-third (target pairs) of the initial EPR pairs are remained.

\section*{Supplementary Information 3: Simulation model}
Similarly to the simulation of traditional QKD, we consider the case without eavesdropper's disturbance.
Here, we consider that the quantum channel is a pure loss model which is similar with BS. The evolution of Fock state $\ket{n}$, coherent state $\ket{\alpha}$ and cat state after passing through the channel can be given by
\begin{equation}
\begin{aligned}\label{}
\ket{n}&\autorightarrow{\textrm{channel}}{\textrm{}}\sum_{m=0}^{n}\sqrt{C_{n}^{m}\eta_{t}^{m}(1-\eta_{t})^{n-m}}\ket{m}_{\textrm{T}}\ket{n-m}_{\textrm{R}}=\ket{\phi(n)},\\
\ket{\alpha}&\autorightarrow{\textrm{channel}}{\textrm{}}\ket{\alpha\sqrt{\eta_{t}}}_{\textrm{T}}\ket{\alpha\sqrt{1-\eta_{t}}}_{\textrm{R}},\\
\ket{\alpha}\pm\ket{-\alpha}&\autorightarrow{\textrm{channel}}{\textrm{}}\ket{\alpha\sqrt{\eta_{t}}}_{\textrm{T}}\ket{\alpha\sqrt{1-\eta_{t}}}_{\textrm{R}}\pm\ket{-\alpha\sqrt{\eta_{t}}}_{\textrm{T}}\ket{-\alpha\sqrt{1-\eta_{t}}}_{\textrm{R}}=\ket{\psi},\\
\end{aligned}
\end{equation}
where $C_{n}^{m}$ is the binomial coefficient and $\eta_{t}$ is the transmittance of channel. The modes $\textrm{T}$ and $\textrm{R}$ will keep in the channel and couple to the environment, respectively. Therefore, the kept quantum states in the channel will be
\begin{equation}
\begin{aligned}\label{}
\rho_{\textrm{T}}(\ket{n})&=\textrm{Tr}_{\textrm{R}}\left(\ket{\phi(n)}\bra{\phi(n)}\right)=\sum_{m=0}^{n}C_{n}^{m}\eta_{t}^{m}(1-\eta_{t})^{n-m}\ket{m}_{\textrm{T}}\bra{m},\\
\rho_{\textrm{T}}(\ket{\alpha})&=\textrm{Tr}_{\textrm{R}}\left(\ket{\alpha\sqrt{\eta_{t}}}_{\textrm{T}}\bra{\alpha\sqrt{\eta_{t}}}\ket{\alpha\sqrt{1-\eta_{t}}}_{\textrm{R}}\bra{\alpha\sqrt{1-\eta_{t}}}\right)=\ket{\alpha\sqrt{\eta_{t}}}_{\textrm{T}}\bra{\alpha\sqrt{\eta_{t}}},\\
\rho_{\textrm{T}}(\ket{\alpha}\pm\ket{-\alpha})&=\textrm{Tr}_{\textrm{R}}\left(\ket{\psi}\bra{\psi}\right)=\ket{\alpha\sqrt{\eta_{t}}}_{\textrm{T}}\bra{\alpha\sqrt{\eta_{t}}}+\ket{-\alpha\sqrt{\eta_{t}}}_{\textrm{T}}\bra{-\alpha\sqrt{\eta_{t}}}\pm e^{-2\mu(1-\eta_{t})}(\ket{\alpha\sqrt{\eta_{t}}}_{\textrm{T}}\bra{-\alpha\sqrt{\eta_{t}}}+\ket{-\alpha\sqrt{\eta_{t}}}_{\textrm{T}}\bra{\alpha\sqrt{\eta_{t}}}),\\
\end{aligned}
\end{equation}
After passing through the channel, the Fock state $\ket{n}$ will become the mixed Fock state with $m$ ($0\leq m \leq n$) photons while the coherent state is still a coherent state containing $\mu\eta_{t}$ photons on average. The detection operation of threshold detector can be characterized by two measurement operators, click $F^{c}$ and no click $F^{nc}$,
\begin{equation}
\begin{aligned}\label{}
F^{c}&=\sum_{n=0}^{\infty}[1-(1-p_{d})(1-\eta_{d})^{n}]\ket{n}\bra{n},\\
F^{nc}&=I-F^{c}=\sum_{n=0}^{\infty}(1-p_{d})(1-\eta_{d})^{n}\ket{n}\bra{n},
\end{aligned}
\end{equation}
where $I=\Sigma_{n=0}^{\infty}\ket{n}\bra{n}$ is the identity operator, $p_{d}$ and $\eta_{d}$ are the dark count rate and efficiency of detector, respectively.

After some calculation, the correct gain $Q_{Z}^{C}$ and error gain $Q_{Z}^{E}$ of the $Z$ basis with coherent state coding can be written as
\begin{equation}
\begin{aligned}\label{}
Q_{Z}^{C}&=(1-p_{d})[1-(1-p_{d})e^{-\mu_{a}\eta_{a}-\mu_{a}\eta_{b}}],\\
Q_{Z}^{E}&=p_{d}(1-p_{d})e^{-\mu_{a}\eta_{a}-\mu_{a}\eta_{b}},\\
\end{aligned}
\end{equation}
where $\mu_{a}\eta_{a}=\mu_{b}\eta_{b}$, $\eta_{a(b)}=\eta_{d}\eta_{at(bt)}$, $\eta_{at(bt)}=10^{-\beta L_{ac(bc)}/10}$, $\beta$ is the the intrinsic loss coefficient of fiber
channel and $L_{ac(bc)}$ is the distance between Alice and Charlie (Bob and Charlie).
Similarly, the correct gain $Q_{X}^{C}$ and error gain $Q_{X}^{E}$ of the $X$ basis with cat state coding can be written as
\begin{equation}
\begin{aligned}\label{}
Q_{X}^{C}&=\frac{1-p_{d}}{2}\Big[1-e^{-2\mu_{a}-2\mu_{b}}-(1-2p_{d})(e^{-\mu_{a}\eta_{a}-\mu_{b}\eta_{b}}-e^{-2\mu_{a}-2\mu_{b}+\mu_{a}\eta_{a}+\mu_{b}\eta_{b}})\Big],\\
Q_{X}^{E}&=\frac{1-p_{d}}{2}\Big[1+e^{-2\mu_{a}-2\mu_{b}}-(1-2p_{d})(e^{-\mu_{a}\eta_{a}-\mu_{b}\eta_{b}}+e^{-2\mu_{a}-2\mu_{b}+\mu_{a}\eta_{a}+\mu_{b}\eta_{b}})\Big].\\
\end{aligned}
\end{equation}
Thereby, the total gain of the $Z$ basis, the QBERs of the $Z$ basis $E_{Z}$ and the $X$ basis $E_{X}$ can be given by
\begin{equation}
\begin{aligned}\label{}
Q_{Z}&=Q_{Z}^{C}+Q_{Z}^{E},\\
E_{Z}&=[e_{d_{Z}}Q_{Z}^{C}+(1-e_{d_{Z}})Q_{Z}^{E}]/Q_{Z},\\
E_{X}&=Q_{X}^{E}/(Q_{X}^{C}+Q_{X}^{E})=Q_{X}^{E}/Q_{Z},
\end{aligned}
\end{equation}
where $e_{d_{Z}}$ is the misalignment rate of the $Z$ basis.

For the protocol in Ref.~\cite{cui:2018:phase1,curty:2018:simple1} with phase-randomized coherent state $\ket{e^{i\theta_{a}}\sqrt{\nu_{a}}}_{a}\ket{e^{i\theta_{b}}\sqrt{\nu_{b}}}_{b}$, the corresponding gain can be given by
\begin{equation}
\begin{aligned}\label{}
Q_{\nu_{a}\nu_{b}}=2(1-p_{d})e^{-\frac{1}{2}(\nu_{a}\eta_{a}+\nu_{b}\eta_{b})}I_{0}(\sqrt{\nu_{a}\eta_{a}\nu_{b}\eta_{b}})-2(1-p_{d})^{2}e^{-(\nu_{a}\eta_{a}+\nu_{b}\eta_{b})},
\end{aligned}
\end{equation}
where $I_{0}(x)$ is the modified Bessel function of the first kind and $I_{0}(0)=1$. The density matrix of the phase-randomized coherent state is
\begin{equation}
\begin{aligned}\label{}
\rho&=\frac{1}{4\pi^{2}}\int_{0}^{2\pi}\int_{0}^{2\pi}\ket{e^{i\theta_{a}}\sqrt{\nu_{a}}}\bra{e^{i\theta_{a}}\sqrt{\nu_{a}}}\ket{e^{i\theta_{b}}\sqrt{\nu_{b}}}\bra{e^{i\theta_{b}}\sqrt{\nu_{b}}}d\theta_{a}d\theta_{b}\\
&=e^{-(\nu_{a}+\nu_{b})}\sum_{n=0}^{\infty}\sum_{m=0}^{\infty}\frac{\nu_{a}^{n}\nu_{b}^{m}}{n!m!}\ket{n}\bra{n}\ket{m}\bra{m},
\end{aligned}
\end{equation}
which is the mixture of Fock states. Let yield $Y_{n,m}$ denote the detection probability when Alice and Bob send Fock states with $n$ and $m$ photons, respectively. Therefore, the gain with intensities $\nu_{a}$ and $\nu_{b}$ can be represented by
\begin{equation}
\begin{aligned}\label{}
Q_{\nu_{a}\nu_{b}}=e^{-(\nu_{a}+\nu_{b})}\sum_{n=0}^{\infty}\sum_{m=0}^{\infty}\frac{\nu_{a}^{n}\nu_{b}^{m}}{n!m!}Y_{n,m}.
\end{aligned}
\end{equation}
The yield $Y_{n,m}$ can be written as
\begin{equation}
\begin{aligned}\label{}
Y_{n,m}=&\sum_{k=0}^{n}\sum_{l=0}^{m}\Bigg\{C_{n}^{k}C_{m}^{l}\eta_{at}^{k}\eta_{bt}^{l}(1-\eta_{at})^{n-k}(1-\eta_{bt})^{m-l}\sum_{u=0}^{k+l}\Bigg\{\frac{u!(k+l-u)!}{2^{k+l}k!l!}\left[\sum_{v=0}^{l}(-1)^{l-v}C_{l}^{v}C_{k}^{u-v}\right]^{2}\\
&\times\left\{[1-(1-p_{d})(1-\eta_{d})^{u}](1-p_{d})(1-\eta_{d})^{k+l-u}+(1-p_{d})(1-\eta_{d})^{u}\left[1-(1-p_{d})(1-\eta_{d})^{k+l-u}\right]\right\}\Bigg\}\Bigg\},
\end{aligned}
\end{equation}
which can be precisely obtained by exploiting the decoy-state method \cite{Hwang:2003:Quantum1,wang2005beating1,lo2005decoy1} with infinite intensities. However, the tight analytical method has been provided in main text by using the three-intensity with $0<\omega<\nu$.



\end{document}